\documentclass[doublecol]{epl2} 

\usepackage{graphicx}
\usepackage{amsmath, amssymb}
\usepackage{lmodern}
\usepackage[T1]{fontenc}
\usepackage{subcaption}
\usepackage[utf8]{inputenc}
\usepackage[english]{babel}
\usepackage{float}
\usepackage{color}
\usepackage{braket}
\usepackage{colortbl}
\usepackage{xcolor}
\title{Adaptive movement strategy may promote biodiversity in the rock-paper-scissors model}

\author{J. Menezes \inst{1,2} \and M. Tenorio \inst{1}  \and E. Rangel\inst{1}}

\institute{   
  \inst{1} Escola de Ci\^encias e Tecnologia, Universidade Federal do Rio Grande do Norte\\
Caixa Postal 1524, 59072-970, Natal, RN, Brazil \\                 
  \inst{2} Institute for Biodiversity and Ecosystem
Dynamics, University of Amsterdam, Science Park 904, 1098 XH
Amsterdam, The Netherlands \\}

\pacs{87.23.-n}{Ecology and evolution}

\abstract{
We study the role of the adaptive movement strategy in promoting
biodiversity in cyclic models described by the rock-paper-scissors game rules. We assume that individuals of one out of the species may adjust their movement to escape hostile regions and stay longer in their comfort zones. Running a series of stochastic simulations, we 
calculate the alterations in the spatial patterns and population densities in scenarios where not all organisms are physically or cognitively conditioned to perform the behavioural strategy.
Although the adaptive movement strategy is not profitable in terms of territorial dominance for the species, it may promote biodiversity. Our findings show that if all individuals are apt to move adaptively, coexistence probability increases for intermediate mobility. The outcomes also show that even if not all individuals can react to the signals received from the neighbourhood, biodiversity is still benefited, but for a shorter mobility range. 
We find that the improvement in the coexistence conditions is more accentuated if organisms adjust their movement intensely and can receive sensory information from longer distances. We also discover that biodiversity is slightly promoted for high mobility if the proportion of individuals participating in the strategy is low. Our results may be helpful for biologists and data scientists to understand adaptive process learning in system biology.}

\begin{document}

\maketitle

\section{Introduction}

Adaptive movement strategies have been observed in many animals that scan and interpret environmental cues, adjusting the speed (kinesis behaviour) or direction (taxis movement) \cite{Motivation1,kinesisnew1,adaptive1,adaptive2,Dispersal,BENHAMOU1989375,coping,NEW}. 
The ability to decipher environmental signals 
is a result of an innate or conditioned behaviour that allows organisms to
avoid enemies' attacks, staying longer in their comfort zone where natural resources are not
scarce \cite{foraging,BUCHHOLZ2007401}. 
Individuals of many species can distinguish areas with an abundance of repellent or attractant species, spending more time in more attractive areas where they are more likely to dominate 
\cite{repellent,Causes,MovementProfitable}.
Motivated by the knowledge of adaptive 
movement behaviour, researchers have improved mechanisms of artificial
intelligence in animats \cite{animats}. Engineers have developed sophisticated engineering tools
driving robots to be aware of the area surrounding them, detecting the presence
of potential threats and smartly reacting either accelerating (decelerating in cases
of detecting a tranquil vicinity) or
moving in a different direction (or keep moving forward if the way is not obstructed).

In this letter, we focus on the adaptive movement strategy performed by
organisms in systems with cyclic dominance.
This a common aspect of species in competition for space in nature,
as observed in relationships among strains of bacteria 
\textit{Escherichia coli},lizards and coral reefs \cite{Coli,bacteria,lizards,Extra1}.
It has been shown that the organisms' adaptive behaviour may promote biodiversity maintenance in cyclic models. For example, local fitness adaptive mobility and mutation may increase the chances of the species persisting increases\cite{mutation0,mutation1}.

We address whether the adaptive movement strategy performed by one of the species is advantageous in terms of population growth and may promote biodiversity in the rock-paper-scissors model.
In general, the standard May-Leonard implementation of the
rock-paper-scissors game considers that organisms of every species move according
to the random walk theory with constant mobility \cite{Reichenbach-N-448-1046, Moura,Anti1,Anti2,Agg,mutation2}. However, some
recent works have shown that taxis-type movement strategy may lead species 
to predominate in the cyclic spatial game \cite{Moura,Agg}. It has also been shown that whether organisms
of one out the species learn to move strategically, biodiversity is affected, mainly being jeopardised \cite{Moura}. We focus on a kinesis-type movement, where
individuals interpret sensory information to adjust
their mobility, either decelerating if in an attractive neighbourhood, or accelerating if
in a hostile area, or keeping constant mobility if the vicinity dangers countervail the advantages of staying there.  

We follow a recent model where the 
neighbourhood attractiveness guides organisms of one species to control their mobility
in the spatial rock-paper-scissors game \cite{Adap}. In this model, the physical ability 
to scan the signals surrounding the organism is modelled by a perception radius, meaning
the maximum distance an organism can perceive the neighbourhood. Furthermore,
the intensity of the reaction to the environment cues is given by a responsiveness strength.
However, it has been reported that not all organisms can execute the species'
behavioural strategy \cite{doi:10.1002/ece3.4446,SabelisII}. Some of them are not physiologically prepared to feel the presence
of repellent and attractant species; others are not capable of associating the 
environment signals to the correct response of adjusting the movement. Because of
this, we sophisticate the model introduced in Ref.~\cite{Adap} by introducing a conditioning factor
that indicates the proportion of organisms that is physically, psychologically, 
and cognitively ready to move adaptively. 

\begin{figure}[t]
	\centering
	\includegraphics[width=40mm]{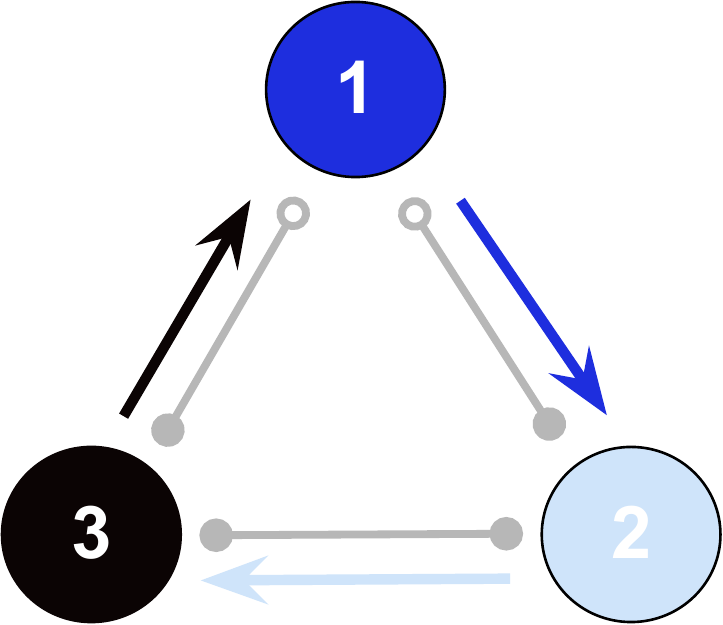}
    \caption{Illustration of the rules in the rock-paper-scissors model. Dark blue, light blue and black arrows show that species $i$ beats species $i+1$. Grey bars represent the mobility interactions; the open circles indicate that individuals of species $1$ can adjust their mobility. 
}
  \label{fig1}
\end{figure}
\begin{figure*}
\centering
    \begin{subfigure}{.23\textwidth}
        \centering
        \includegraphics[width=40mm]{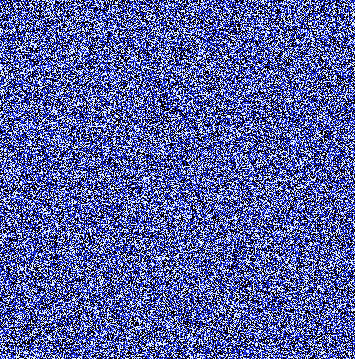}
        \caption{}\label{fig2a}
    \end{subfigure} %
    \begin{subfigure}{.23\textwidth}
        \centering
        \includegraphics[width=40mm]{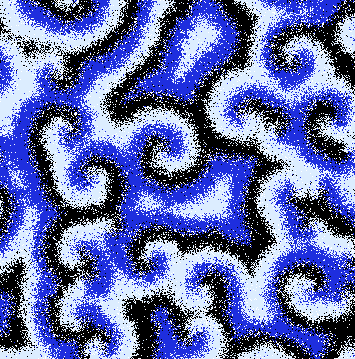}
        \caption{}\label{fig2b}
    \end{subfigure} %
    \begin{subfigure}{.23\textwidth}
        \centering
        \includegraphics[width=40mm]{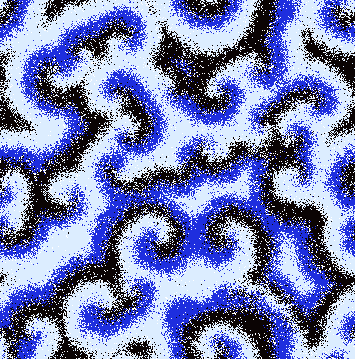}
        \caption{}\label{fig2c}
    \end{subfigure} %
          \begin{subfigure}{.23\textwidth}
        \centering
        \includegraphics[width=40mm]{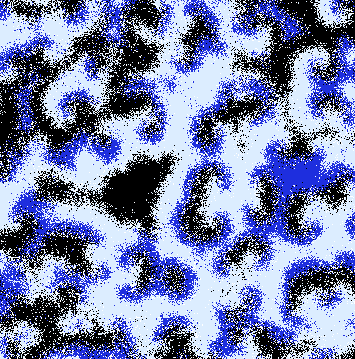}
        \caption{}\label{fig2d}
    \end{subfigure} %
    \caption{Spatial patterns captured from simulations of the rock-paper-scissors game illustrated in Fig.~\ref{fig1} running in grids with $500^2$ sites. Figure \ref{fig2a} shows the initial conditions used to run the simulations whose spatial configuration at $t=5000$ generations are shown in Fig.~\ref{fig2b} ($\alpha=0.0$), ~\ref{fig2c} ($\alpha=0.3$), and ~\ref{fig2d} ($\alpha=1.0$).}
  \label{fig2}
\end{figure*}
\begin{figure}
     \centering
    \includegraphics[width=78mm]{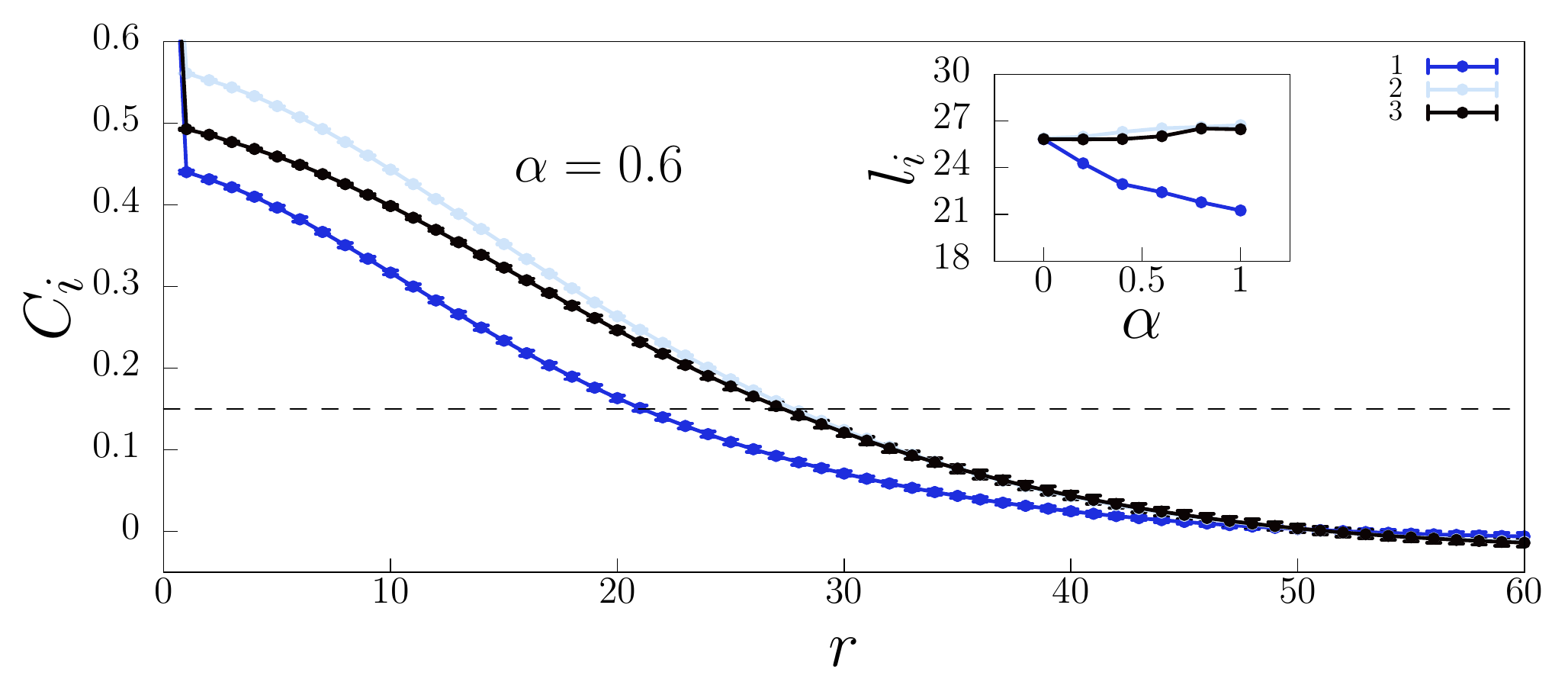}
    \caption{The main figure shows $C_i$ for $\alpha=0.6$; the error bars indicate the standard deviation. The inset figure depicts $l_i$ as a function of $\alpha$; the horizontal dashed black line shows the threshold $C_i = 0.15$.}
  \label{fig3}
\end{figure}


\section{The model}
\label{sec2}

We investigate a cyclic nonhierarchical system composed of three species, whose dominance is described by the rock-paper-scissors game rules. 
We model the behavioural strategy that leads individuals of one out of the species to adjust their mobility to adapt to escape dangerous regions and stay longer in attractive areas. 
We implement stochastic simulations following a standard algorithm widely used to investigate spatial biological systems \cite{Reichenbach-N-448-1046,uneven}, where organisms interact in square lattices with periodic boundary conditions without any conservation law for the total number of individuals (May-Leonard numerical implementation \cite{leonard})
We assume that each grid point contains at most one individual; thus, the maximum number of individuals is $\mathcal{N}$, the total number of grid points.
The selection interactions are illustrated in Fig.~\ref{fig1}, where organisms of species $i$ beat individuals of species $i+1$, with $i=1,2,3$, with the cyclic identification $i=i\pm3\,k$, where $k$ is an integer. The dark blue, light blue, and black arrows represent the cyclic selection dominance of species $1$, $2$, and $3$, respectively. The grey bars represent the position exchange during mobility interactions, with the open circle indicating that only organisms of species $1$ can adjust their mobility to local cues.

All simulations begin from random initial conditions: we
distribute each individual at a random grid point. Initially, 
the number of individuals is the same for all species, i.e., $I_i\,=\,\mathcal{N}/3$, with $i=1,2,3$.
At each timestep, the spatial configuration changes by the implementation of one of the spatial interactions: a) Selection: $ i\ j \to i\ \otimes\,$, with $ j = i+1$, where $\otimes$ means an empty space; b) Reproduction: $ i\ \otimes \to i\ i\,$; c) Mobility: $ i\ \odot \to \odot\ i\,$, where $\odot$ means either an individual of any species or an empty site. 
Whenever one selection interaction occurs, the grid point occupied by the
the eliminated organism becomes an empty space that an offspring of any species may subsequently fill.
The interactions are implemented following the Moore neighbourhood, where individuals may interact with one of their eight nearest neighbours, following the steps: i) randomly choosing an active individual; ii) drawing one interaction to be implemented; iii) raffling one of the eight nearest neighbours to be the passive of the sorted interaction. If the interaction is executed, one timestep is counted. Otherwise, the three steps are repeated. We define the time unit as the time necessary to $\mathcal{N}$ timesteps to occur: one generation.

For every species, selection and reproduction interactions are implemented following the probabilities $s$ and $r$. In addition, $m$ is the maximum mobility probability an organism of species $1$ can move, so that $s+r+m=1$. However, individuals of species $2$ and $3$ move 
with constant mobility probability given by
$m_2\,=\,m_3\,=\,m/2$. For individuals of species $1$, the mobility probability, $m_1$ is adjusted according to the local species densities so that, $m_1\,<\,m/2$ in attractive zones and $m_1\,>\,m/2$ in hostile regions.

The adaptive movement behaviour of organisms of species $1$ follows the algorithm introduced in Ref.~\cite{Adap} with a conditioning factor $\alpha$, a real parameter, with $0.0 \leq \alpha \leq 1.0$. This parameter indicates the probability of the individual being able to adjust its movement in response to local stimuli.
We also assume that the maximum Euclidian distance an organism can perceive the environmental signals is $R$, the perception radius ($R$ is an integer measured in lattice spacing).
We compute
the local densities of dominant (species $3$) and dominated (species $2$) individuals, $d_3$ and $d_2$, respectively, within a disk of radius $R$ around the active individual. Based on the neighbourhood scan, 
an organism of species $1$ calculates local attractiveness, $\mathcal{A}$ as the difference between the local densities of individuals of species $2$ and $3$; thus, adjusting the mobility within the range 
$0\,<\,m_1\,<\,m$, employing the function

\begin{equation}
m_1\,=\, \,\frac{m}2\,\Big[1\,+\,\tanh \left(-\epsilon\,\mathcal{A}\right)\Big],
\end{equation}

where $\epsilon$ is a real parameter representing the organism's responsiveness to the local stimuli. The local density of conspecifics does not influence the organism's decision-making for changing its mobility.

\begin{figure}[t]
	\centering
        \begin{subfigure}{.49\textwidth}
        \centering
        \includegraphics[width=75mm]{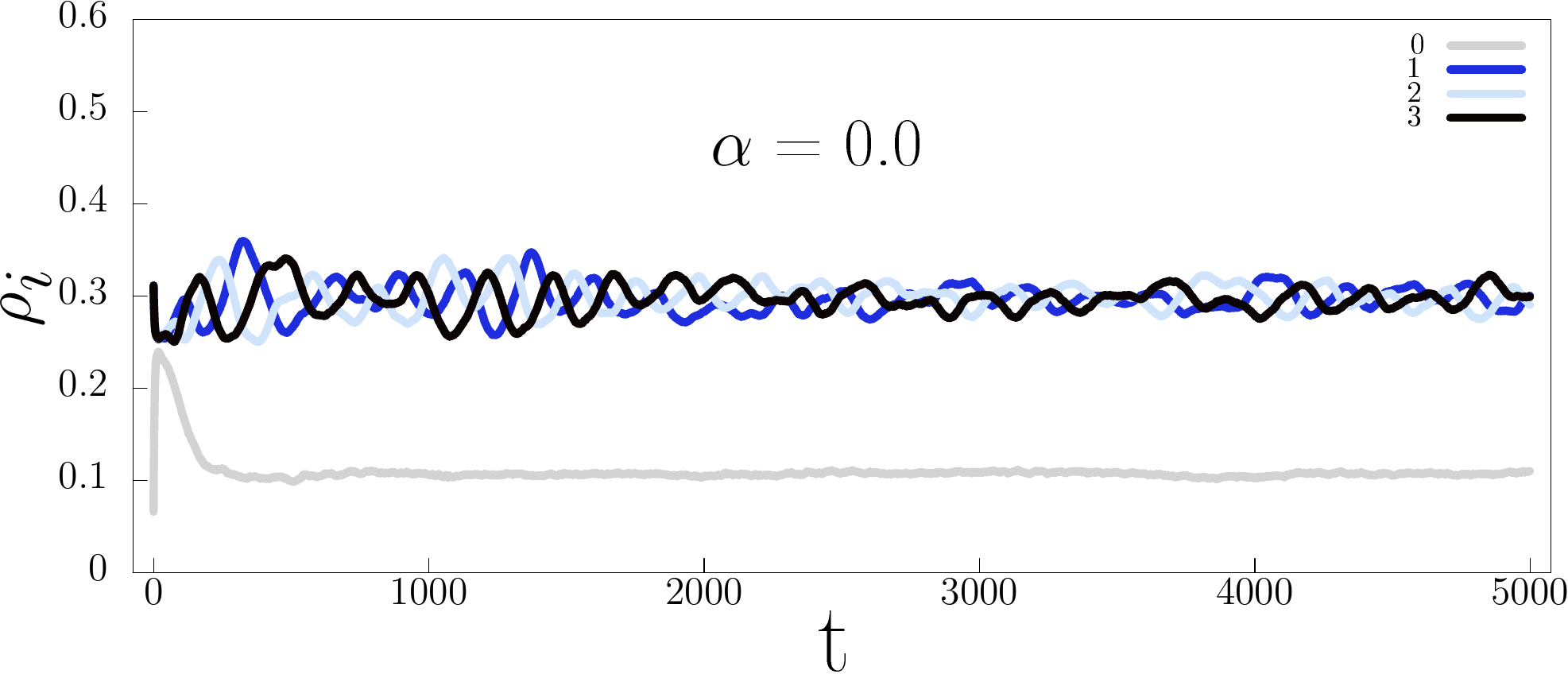}
        \caption{}\label{fig4a}
    \end{subfigure}\\
       \begin{subfigure}{.49\textwidth}
        \centering
        \includegraphics[width=75mm]{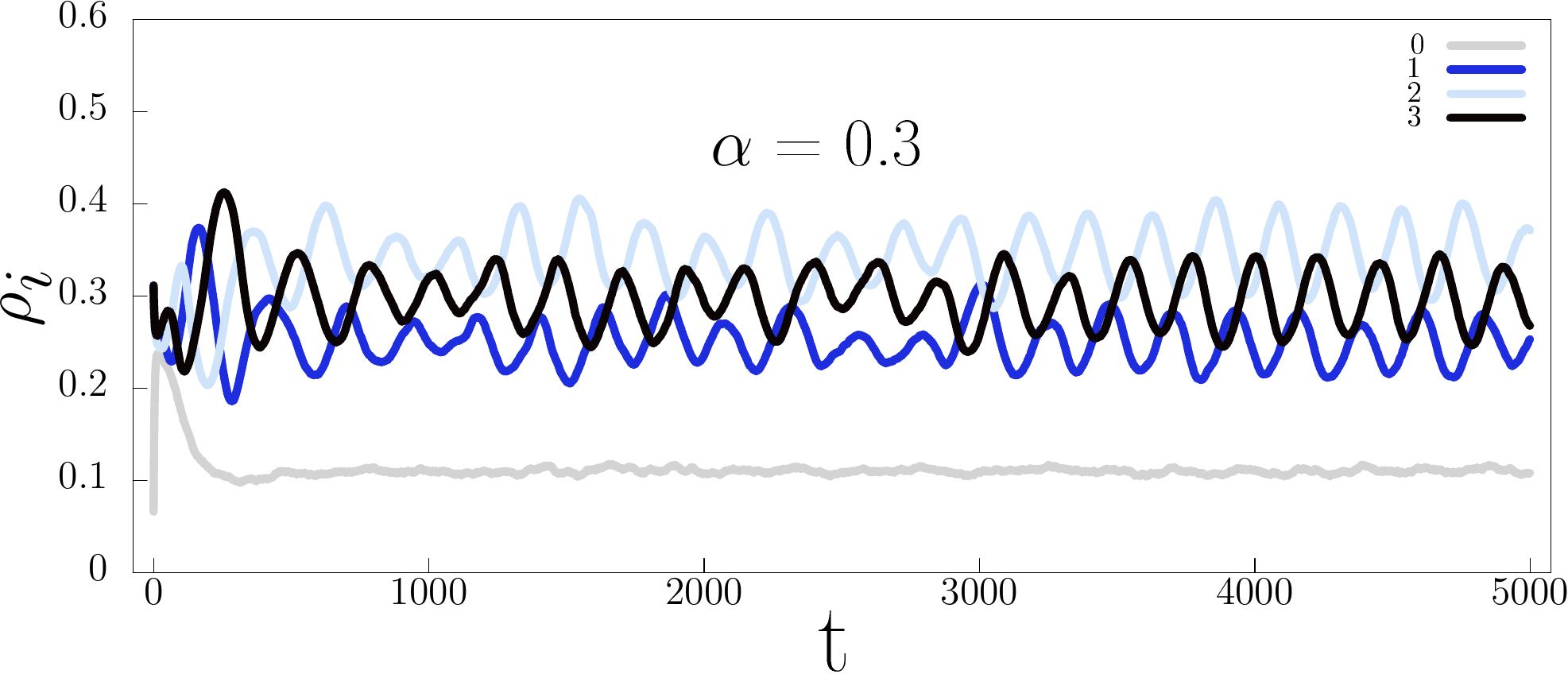}
        \caption{}\label{fig4b}
    \end{subfigure}\\
           \begin{subfigure}{.49\textwidth}
        \centering
        \includegraphics[width=75mm]{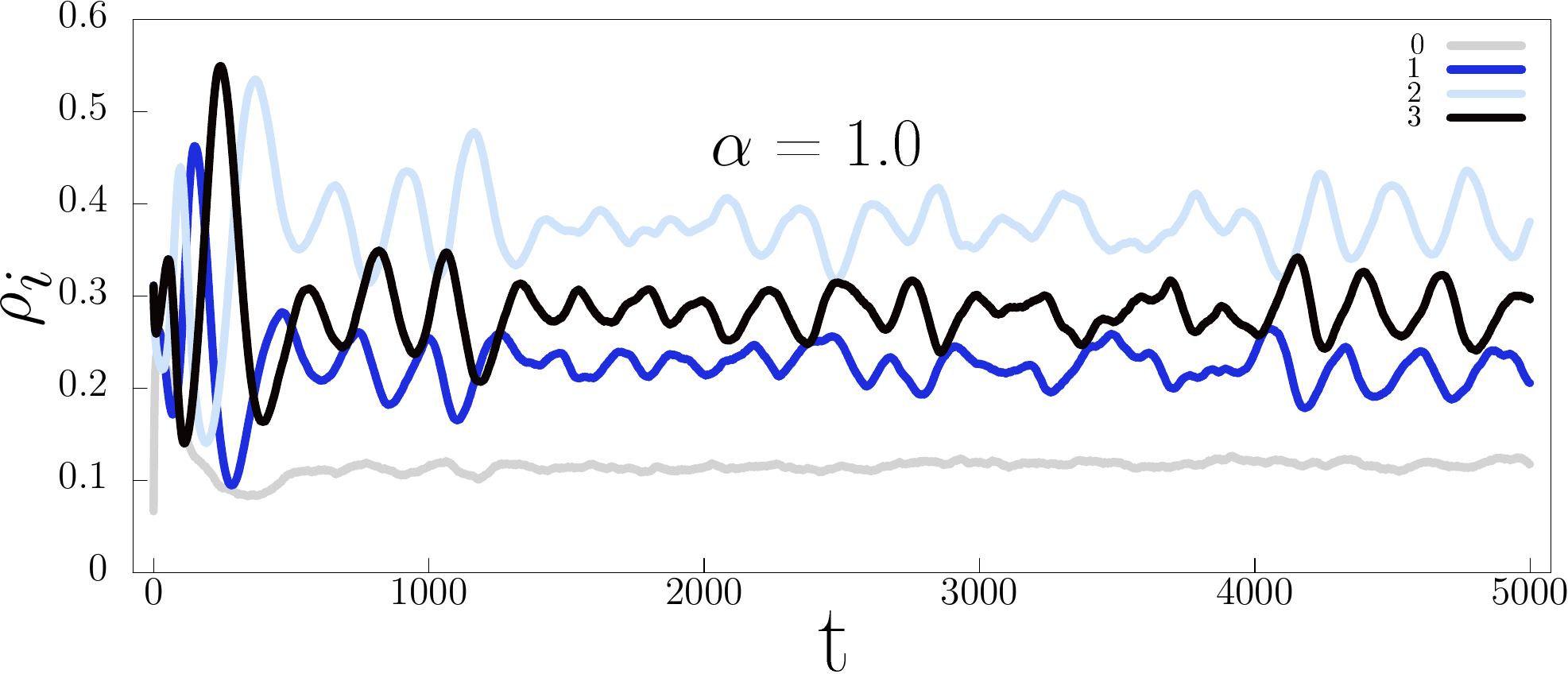}
        \caption{}\label{fig4c}
    \end{subfigure}
 \caption{
Dynamics of the species densities. Figures \ref{fig4a}, \ref{fig4b}, and \ref{fig4c}
show $\rho_i$ as a function of the time for the simulations in Figs. \ref{fig2b}, \ref{fig2c}, and \ref{fig2d}. }
  \label{fig4}
\end{figure}

\section{Spatial Patterns}
\label{sec3}
To study the spatial patterns,
we performed a single realisation for the $\alpha=0.0$ (the standard model), $\alpha=0.3$, and $\alpha=1.0$; the simulations started from the random initial conditions presented in Fig.~\ref{fig2a}.
The realisations were performed in square lattices with $500^2$ grid points for a timespan of $5000$ generations. The perception radius was set $R=2$, while $\epsilon=2$, $s\,=\,r\,=\,0.1$ and $m = 0.8$. 
We captured $500$ snapshots of the lattice (in intervals of $10$ generations); then, we used the snapshots to produce the videos in https://youtu.be/iousy7q-m3s ($\alpha=0.0$), https://youtu.be/bJvZTdidED8 ($\alpha=0.3$), and https://youtu.be/dxivAmeVrBY ($\alpha=1.0$). Figures \ref{fig2b}, ~\ref{fig2c}, and \ref{fig2d} depict the snapshots at the end of the simulations, where organisms of species $1$, $2$, and $3$, are showed by dark blue, light blue, and black dots, respectively; empty spaces are represented by white dots. 

Since the realisations started from random initial conditions, selection interactions are frequent in the initial stage of the simulations. As time passes, single-species spatial domains arise in the form of spirals. Because organisms of every species have constant mobility
in the standard model ($\alpha=0.0$), the formation and dynamics of the spiral patterns are symmetric (Fig.~\ref{fig2b}). However, the spatial configuration is altered if individuals of species $1$ can scan the environment to produce the appropriate motor response. The irregular spiral formation (Fig.~\ref{fig2c} and Fig.~\ref{fig2d}) reflects the turbulence on the spatial segregation of species when individuals of species $1$ no longer move with constant mobility probability. 
The fact that the organisms in the spiral arm of species $1$ move with different velocities unbalances the spiral arms dislocations. Namely, 
individuals in the front have lower mobility probabilities than organisms in the back. Consequently, the spiral arm of species $1$ (dark blue) is narrower than both other arms.
\section{Autocorrelation function}
\label{sec4}
Let us now quantify how the scale of the single-species spatial domains change if the fraction of organisms conditioned to move adaptively grows. For this purpose, we calculate how individuals of the same species are spatially correlated by computing the spatial autocorrelation function $C_i(r)$, with $i=1,2,3$, in terms of radial coordinate $r$, where $r=|\vec{r}|=x+y$ is the Manhattan distance between $(x,y)$ and $(0,0)$. 

In this sense, we introduce the function $\phi_i(\vec{r})$ to represent the presence of an organism of species $i$ in the position $\vec{r}$ in the lattice. We compute the Fourier transform
\begin{equation}
\varphi_i(\vec{\kappa}) = \mathcal{F}\,\{\phi_i(\vec{r})-\langle\phi_i\rangle\},
\end{equation}
where $\langle\phi_i\rangle$ is the mean value of $\phi_i$. After that,
we calculate the spectral densities
$S_i(\vec{k}) = \sum_{k_x, k_y}\,\varphi_i(\vec{\kappa})$.
The autocorrelation function is described by the normalised inverse Fourier transform
\begin{equation}
C_i(\vec{r}') = \frac{\mathcal{F}^{-1}\{S_i(\vec{k})\}}{C(0)},
\end{equation}
that can be written as 
a function of $r$ as
\begin{equation}
C_i(r') = \sum_{|\vec{r}'|=x+y} \frac{C_i(\vec{r}')}{min\left[2N-(x+y+1), (x+y+1)\right]}.
\end{equation}

Figure~\ref{fig3} depict $C_i$ as a function of the radial coordinate $r$ for various values of the conditioning factor. The colours follow the scheme in Fig.~\ref{fig1}, where dark blue, light blue, and black dots indicate the mean values for the species $1$, $2$, and $3$, respectively.
The mean autocorrelation function was averaged from outcomes obtained from a set of $100$ simulations in lattices with $300^2$ grid points,
starting from different initial conditions. The parameters are the same as in
the simulations in Fig.~\ref{fig2}, with $\alpha=0.6$.

To calculate the typical size of the single-species regions, we consider the threshold $C_i(l_i)=0.15$, where $l_i$ is the characteristic length scale for spatial domains of species $i$ (the horizontal black line in Fig.~\ref{fig3}); the inset figure depicts $l_i$. As $\alpha$ grows,
the discrepancy in the typical size of regions controlled by each species
becomes more relevant. The characteristic length for areas inhabited by species organisms $1$ decreases as $\alpha$ grows. On the other hand, $l_2$ elongates, independent of the proportion of individuals of species $1$ conditioned to follow the behavioural strategy. Athough $l_3$ shortens for $\alpha=0.2$, there is an growth of the spatial domains of species $3$ for $\alpha>0.2$, with $l_3<l_2$.
\begin{figure}[t]
\centering
    \includegraphics[width=70mm]{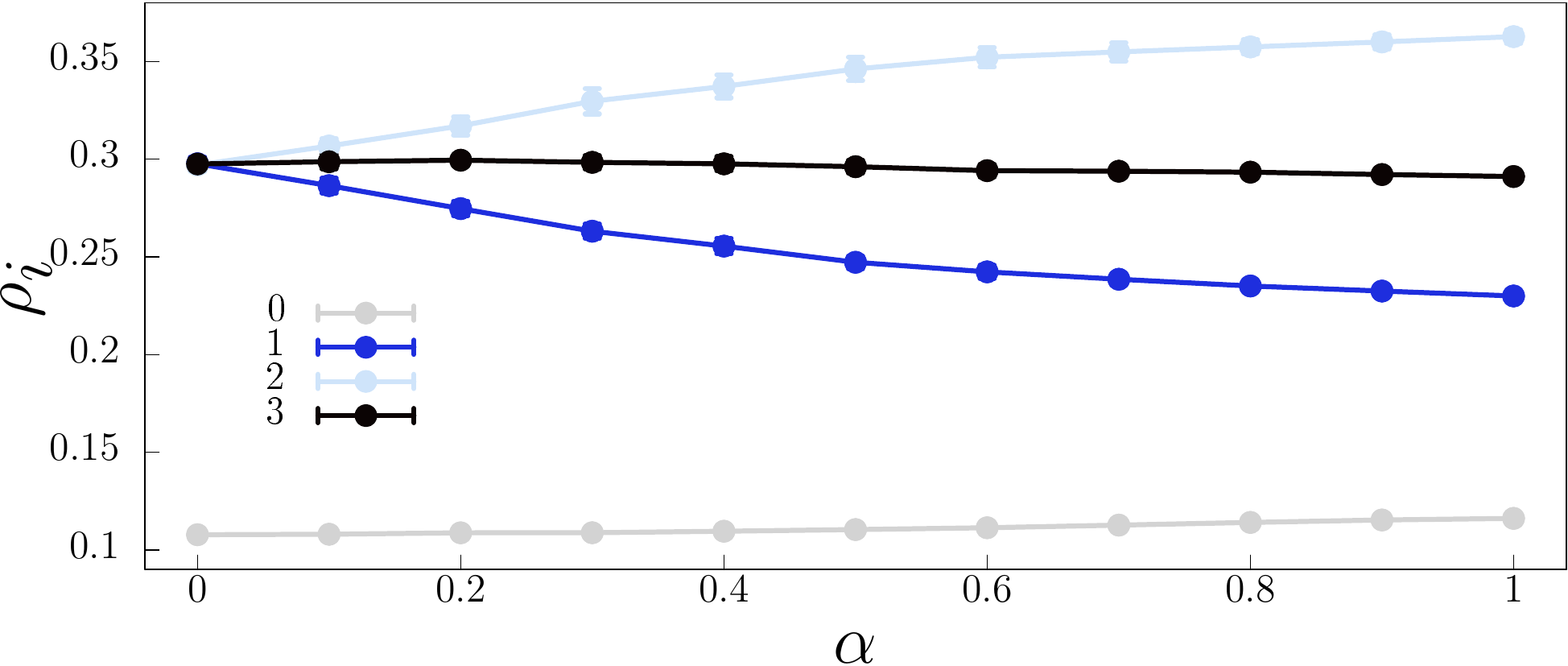}
    \caption{Mean species densities as functions of the conditioning factor. Error bars show the standard deviation.}
  \label{fig5}
\end{figure}
\begin{figure*}[t]
	 \centering
	    \begin{subfigure}{.27\textwidth}
    \centering
    \includegraphics[width=37mm]{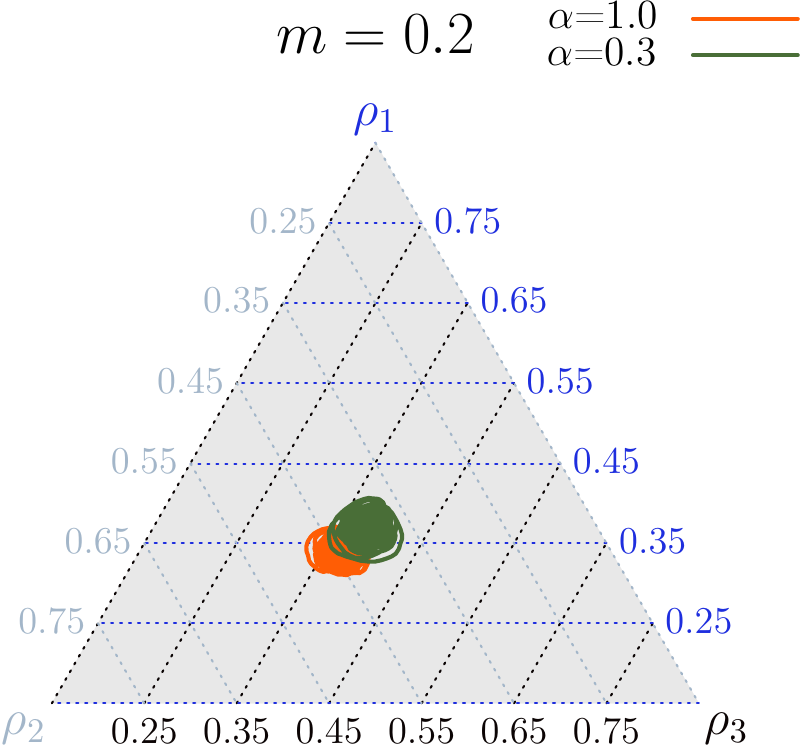}
    \caption{}\label{fig6a}
  \end{subfigure}
    \begin{subfigure}{.27\textwidth}
    \centering
    \includegraphics[width=37mm]{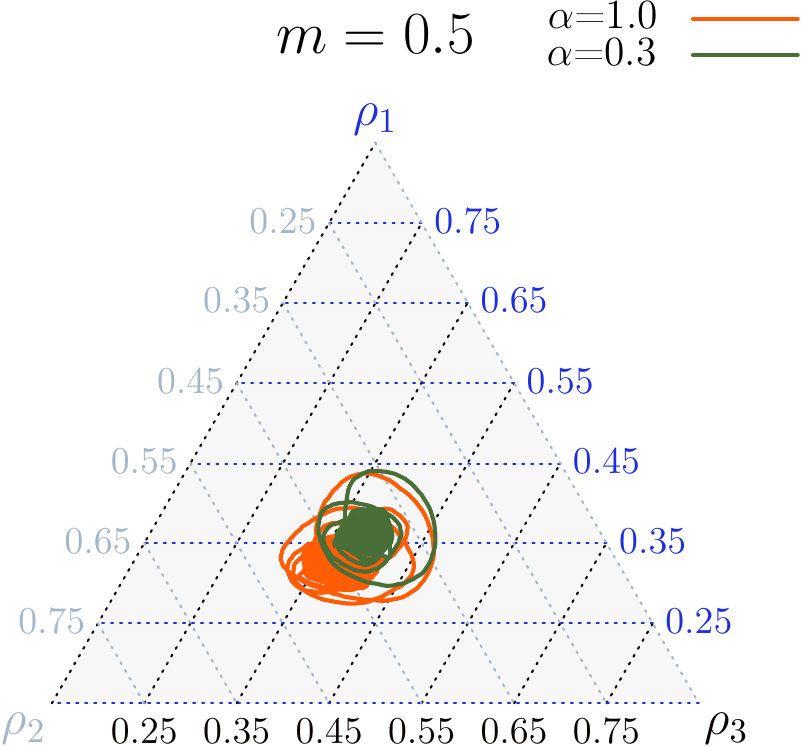}
    \caption{}\label{fig6b}
  \end{subfigure}
      \begin{subfigure}{.27\textwidth}
    \centering
    \includegraphics[width=37mm]{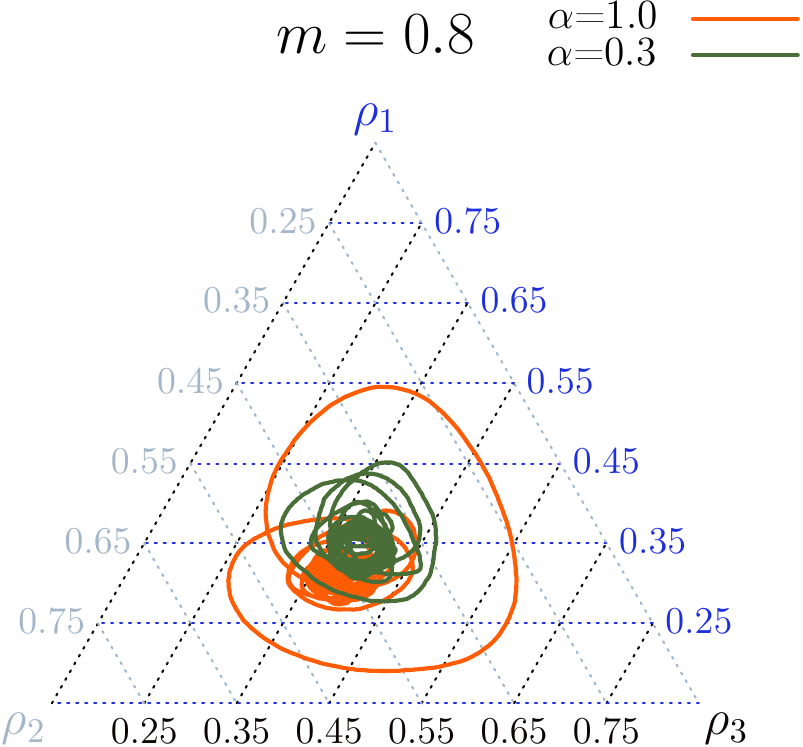}
    \caption{}\label{fig6c}
  \end{subfigure}\\	
    \begin{subfigure}{.27\textwidth}
    \centering
    \includegraphics[width=37mm]{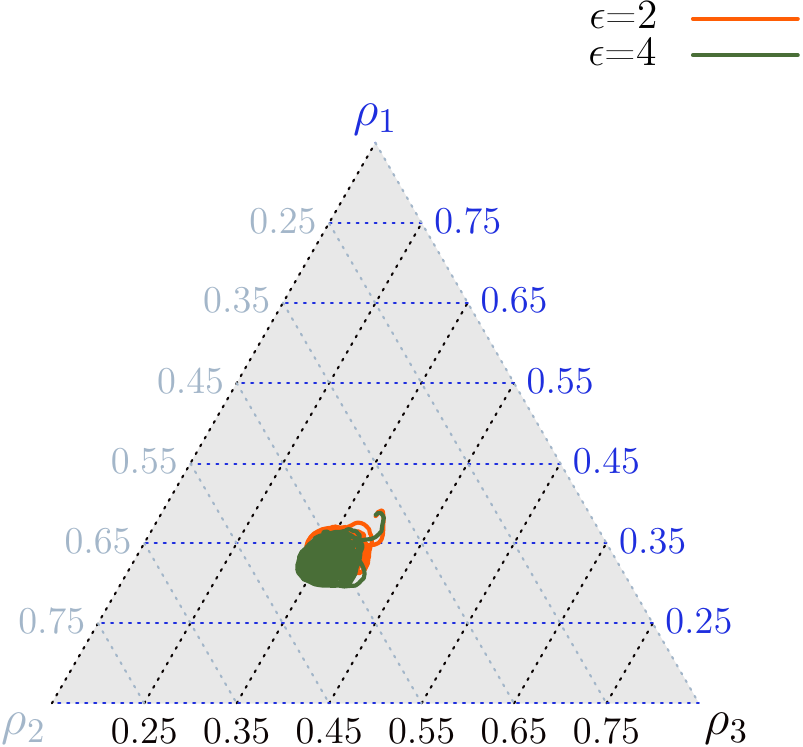}
    \caption{}\label{fig6d}
  \end{subfigure}
    \begin{subfigure}{.27\textwidth}
    \centering
    \includegraphics[width=37mm]{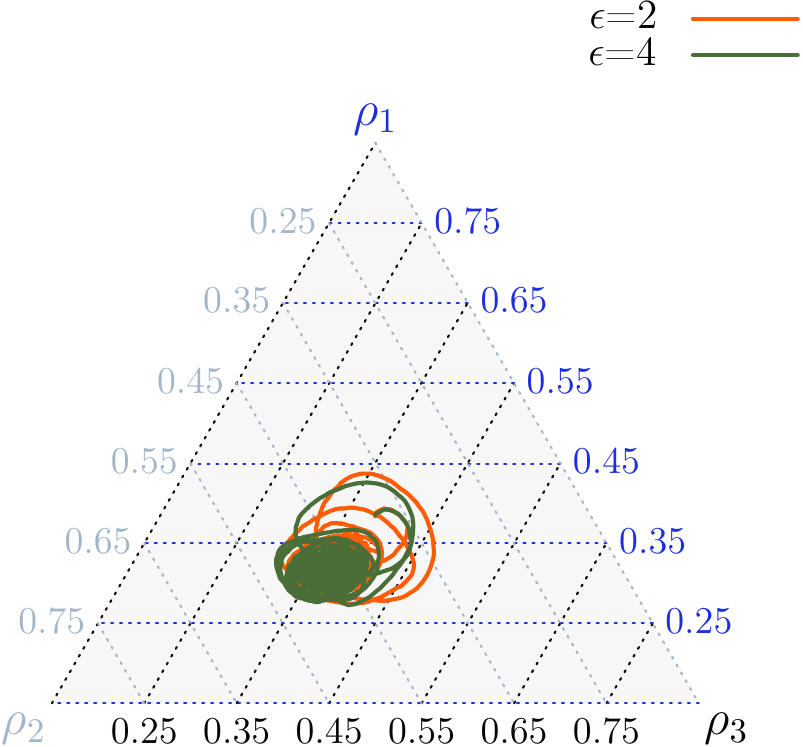}
    \caption{}\label{fig6e}
  \end{subfigure}
      \begin{subfigure}{.27\textwidth}
    \centering
    \includegraphics[width=37mm]{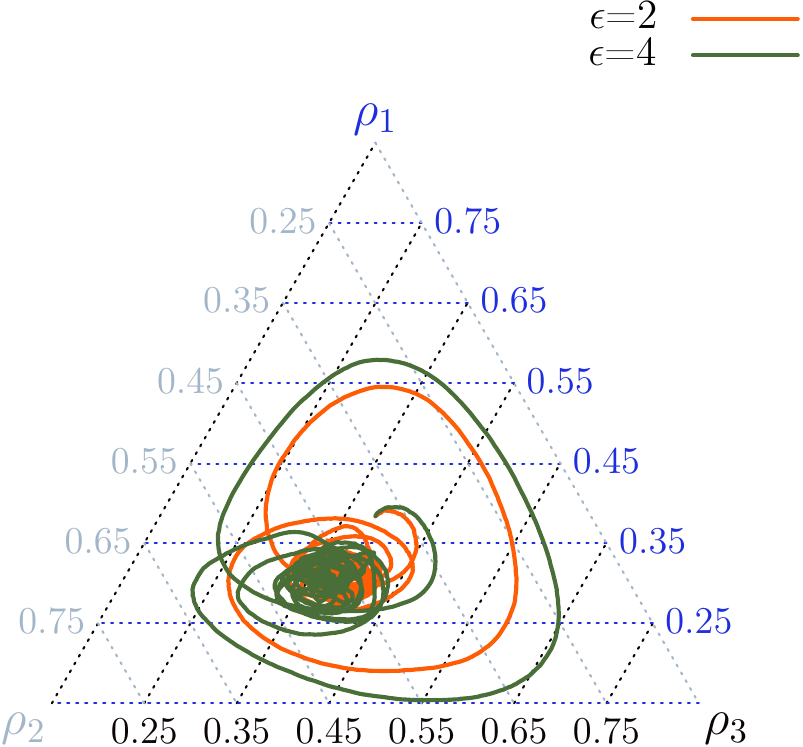}
    \caption{}\label{fig6f}
  \end{subfigure}
  \\
   \begin{subfigure}{.27\textwidth}
    \centering
    \includegraphics[width=37mm]{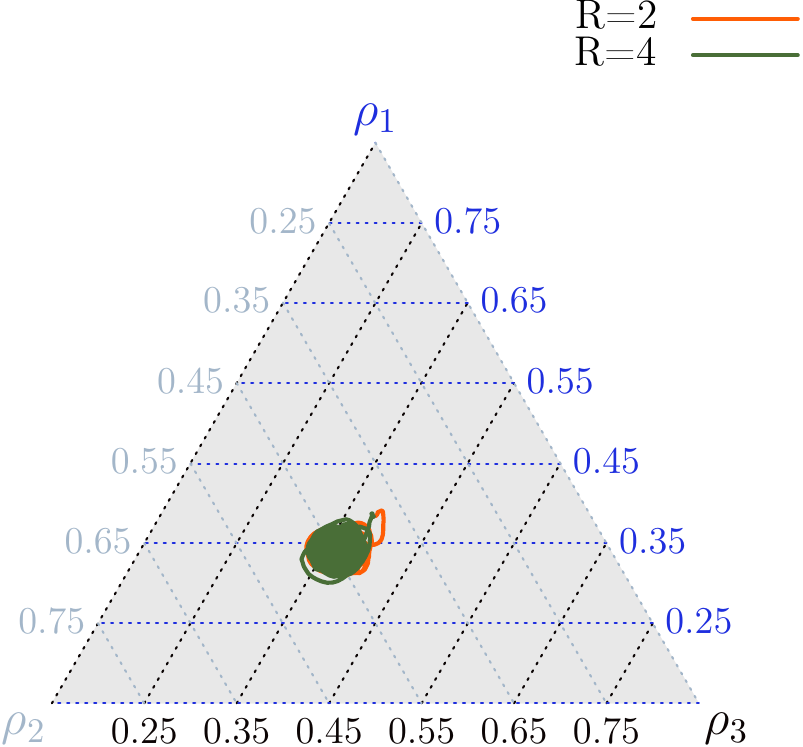}
    \caption{}\label{fig6g}
  \end{subfigure}
    \begin{subfigure}{.27\textwidth}
    \centering
    \includegraphics[width=37mm]{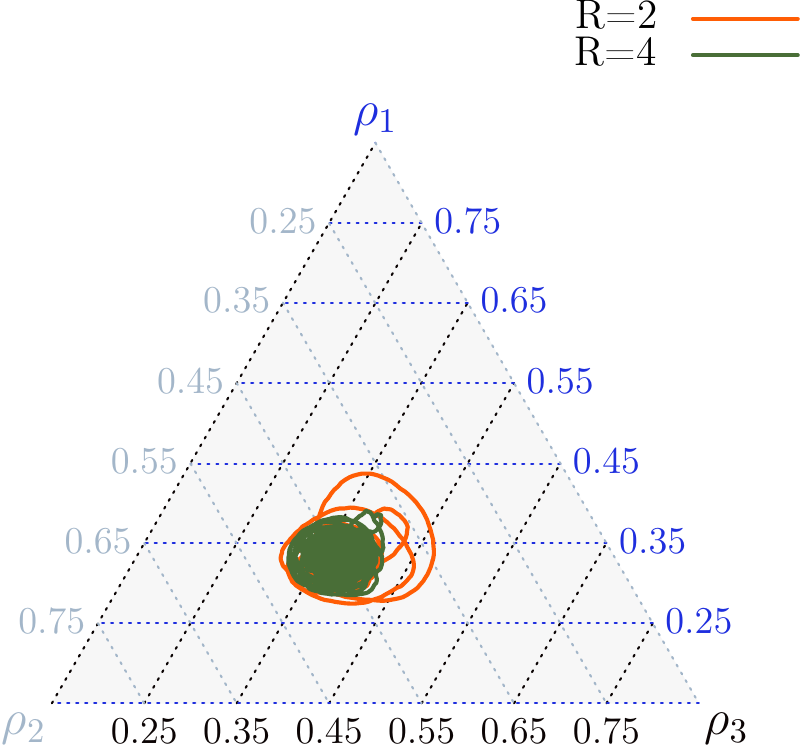}
    \caption{}\label{fig6h}
  \end{subfigure}
      \begin{subfigure}{.27\textwidth}
    \centering
    \includegraphics[width=37mm]{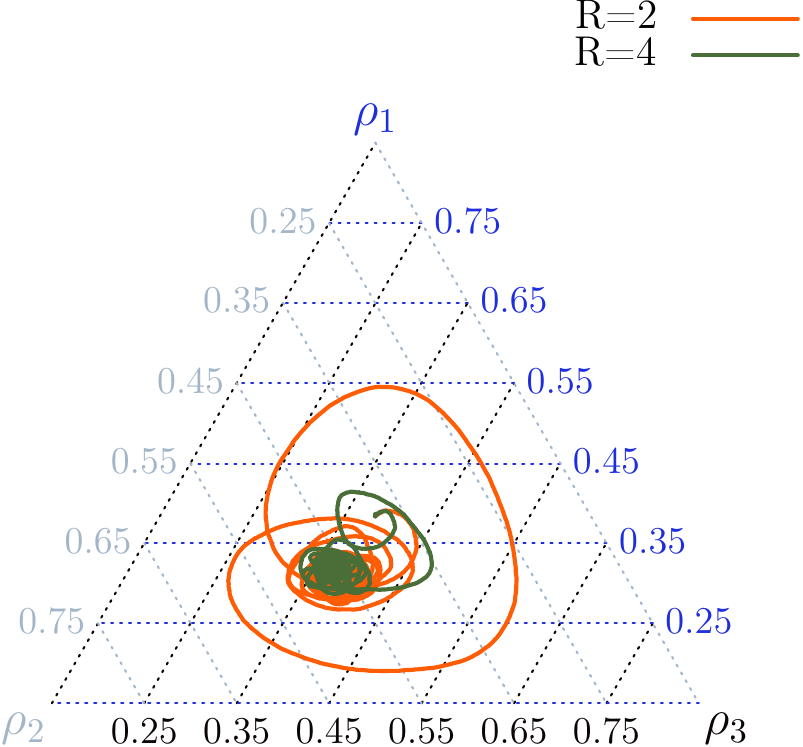}
    \caption{}\label{fig6i}
  \end{subfigure}
    \caption{Ternary diagrams of the dynamics of species densities for various mobility probabilities. The outcomes were obtained for $m=0.2$ (Figs.~\ref{fig6a}, ~\ref{fig6d}, and ~\ref{fig6g}) $m=0.5$ (Figs.~\ref{fig6b}, ~\ref{fig6e}, and ~\ref{fig6h}), and $m=0.8$ (Figs.~\ref{fig6c}, ~\ref{fig6f}, and ~\ref{fig6i}), in lattices with $500^2$ grid points.}
  \label{fig6}
\end{figure*}

\section{Species Densities}
\label{sec5}
We explore the effects of the adaptive movement strategy on the population dynamics by investigating the role of the conditioning factor in the species densities, $\rho_i$, with $i=1,2,3$, defined as the fraction of the grid occupied by individuals of species $i$ at time $t$: 
$\rho_i\,=\,I_i/\mathcal{N}$.
The proportion of empty space is denoted by $\rho_0$.
Figure~\ref{fig4a}, \ref{fig4b}, and \ref{fig4c} depicts the fraction of the grid occupied by each species
in the simulations presented in Fig.~\ref{fig2b}, \ref{fig2c}, and \ref{fig2d}. The grey colour depicts the density of empty space ($\rho_0$) while dark blue, light blue, and black show the results for species $1$, $2$, and $3$, respectively. 

The outcomes reveal that the more individuals of species $1$ learn to perform the adaptive movement strategy, the more
disadvantageous for species $1$ in terms of territorial dominance is. 
The outcomes also indicate that the difference in the species densities increases if $\alpha$ grows, with species $2$ being predominant \cite{Adap}. Moreover, Fig.~\ref{fig4b} shows that the amplitude of the cyclic function $\rho_i$ grows in comparison to the standard model (Fig.~\ref{fig4a}), even if few organisms are ready to move adaptively. This is a result of the irregular spiral pattern formation in Figs.~\ref{fig2b} and~\ref{fig2c}.

\subsection{Mean species densities}

We study the impact of the adaptive movement strategy in the species densities. 
Performing a series of $100$ simulations starting from different initial conditions and running until $5000$ generations, we computed the mean value of the species densities, $\rho_i$ using the parameters $s\,=\,r\,=\,0.1$, $m\,=\,0.8$, $R=2$, and $\epsilon=2$.
Aiming to avoid the density fluctuations in the pattern formation process, we averaged the results considering the numerical data of the second half of the simulations.

Figure~\ref{fig5} shows how species densities vary as more organisms learn to move adaptively. Overall, we found evidence that the larger the number of organisms moving according to local stimulus, the more species populations are affected. Namely, as $\alpha$ grows, the territorial dominance of species $2$ is enhanced, with negative variation for the population of species $1$, whose organisms move intelligently. In the case of species $3$, the slight population growth for $\alpha \leq 0.2$ is substituted for a population decline as more than $20\%$ of individuals of species $1$ apt to adapt their movement.

\subsection{Species densities' oscillations}

The outcomes presented in Fig.~\ref{fig4} revealed that the species densities' fluctuations in the initial simulation stage are accentuated as $\alpha$ grows. For this reason, we aim to discover how the oscillations in the species densities are impacted by the adaptive movement of individuals of species $1$ for different values of organism's responsiveness, perception radius, and conditioning factor.
For this, we considered single simulations for a low ($m=0.2$), intermediate ($m=0.5$), and high ($m=0.8$) mobility; the interaction probabilities are constrainted by $s=r=(1-m)/2$.

The outcomes are shown in Fig.~\ref{fig6}: i) Dark grey ternary diagrams in Figs.~\ref{fig6a}, ~\ref{fig6d}, and ~\ref{fig6g} depict the orbits for $m=0.2$ for $\alpha=0.3$ and $\alpha=1.0$, $\epsilon=2$ and $\epsilon=4$, $R=2$ and $R=4$, respectively; ii) Light grey ternary diagrams in Figs.~\ref{fig6b}, ~\ref{fig6e}, and ~\ref{fig6h} show the orbits for $m=0.5$ for $\alpha=0.3$ and $\alpha=1.0$, $\epsilon=2$ and $\epsilon=4$, $R=2$ and $R=4$, respectively; iii) White ternary diagrams in Figs.~\ref{fig6c}, ~\ref{fig6f}, and ~\ref{fig6i} show the orbits for $m=0.8$ for $\alpha=0.3$ and $\alpha=1.0$, $\epsilon=2$ and $\epsilon=4$, $R=2$ and $R=4$, respectively.

Generally speaking, the species densities' oscillations increases with $m$: if mobility increase, the role of the adaptive movement becomes more significant, thus provoking a more accentuated unevenness in the spatial pattern formation stage. Furthermore,
our discoveries counter-intuitively show that, for intermediate mobility, the oscillations are more accentuated in the case of fewer organisms participating in the adaptive movement tactic (Fig.~\ref{fig6b}), or if the responsiveness to the local stimuli is less intense (Fig.~\ref{fig6e}). This is no longer valid for high mobility, where the more relevant fluctuations in the species' densities are provoked for higher $\alpha$ (Fig.~\ref{fig6c}) and $\epsilon$ (Fig.~\ref{fig6f}). However, for a fixed $\alpha$ and $\epsilon$, the outcomes indicate that the species densities' fluctuations increase if the adaptive movement is performed for individuals with longer range environment perception $R$, irrespective of $m$.

\section{Coexistence probability}
\label{sec6}
\begin{figure}[t]
	\centering
        \begin{subfigure}{.49\textwidth}
        \centering
        \includegraphics[width=78mm]{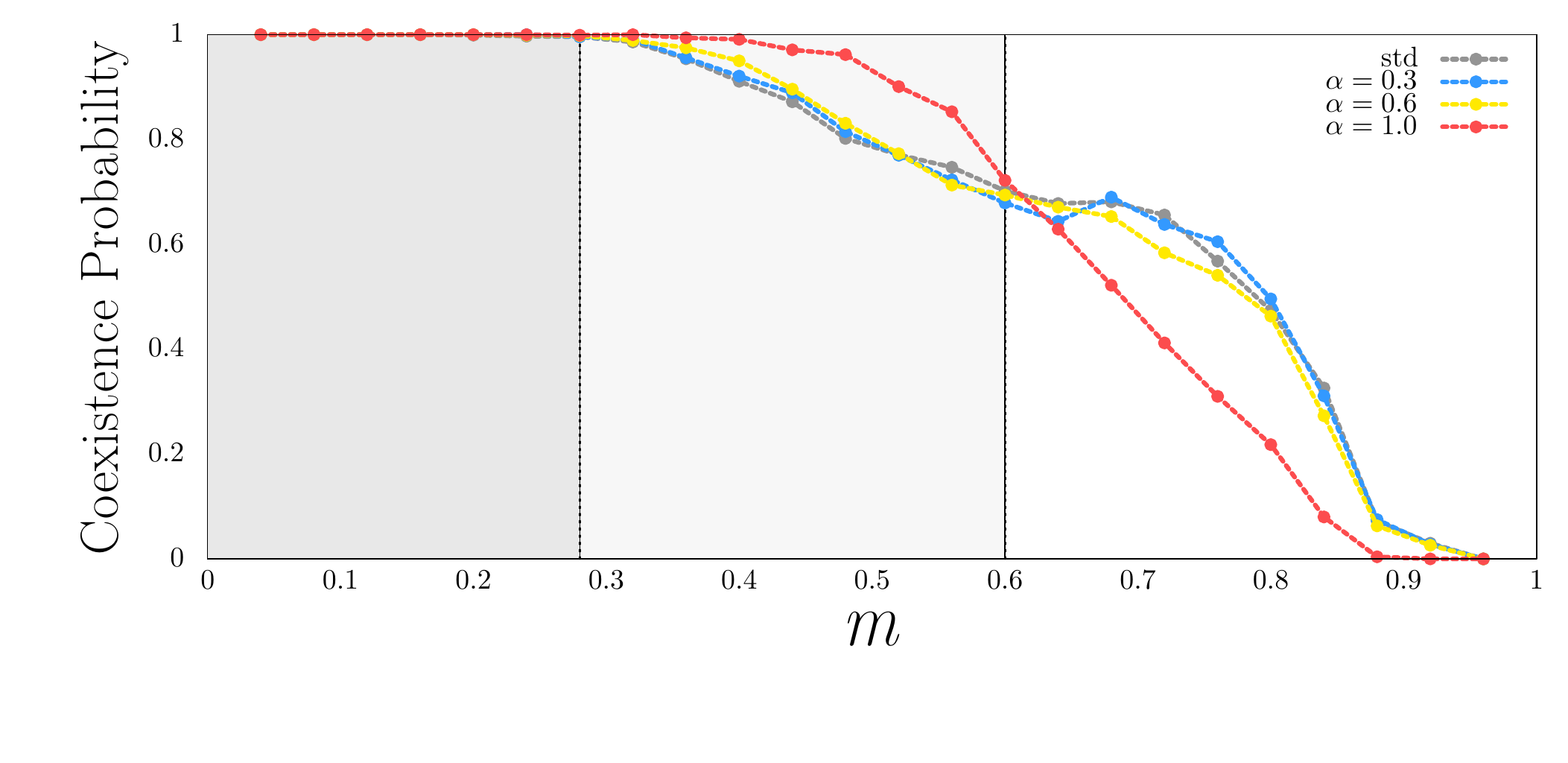}
        \caption{}\label{fig7a}
    \end{subfigure}\\
       \begin{subfigure}{.49\textwidth}
        \centering
        \includegraphics[width=78mm]{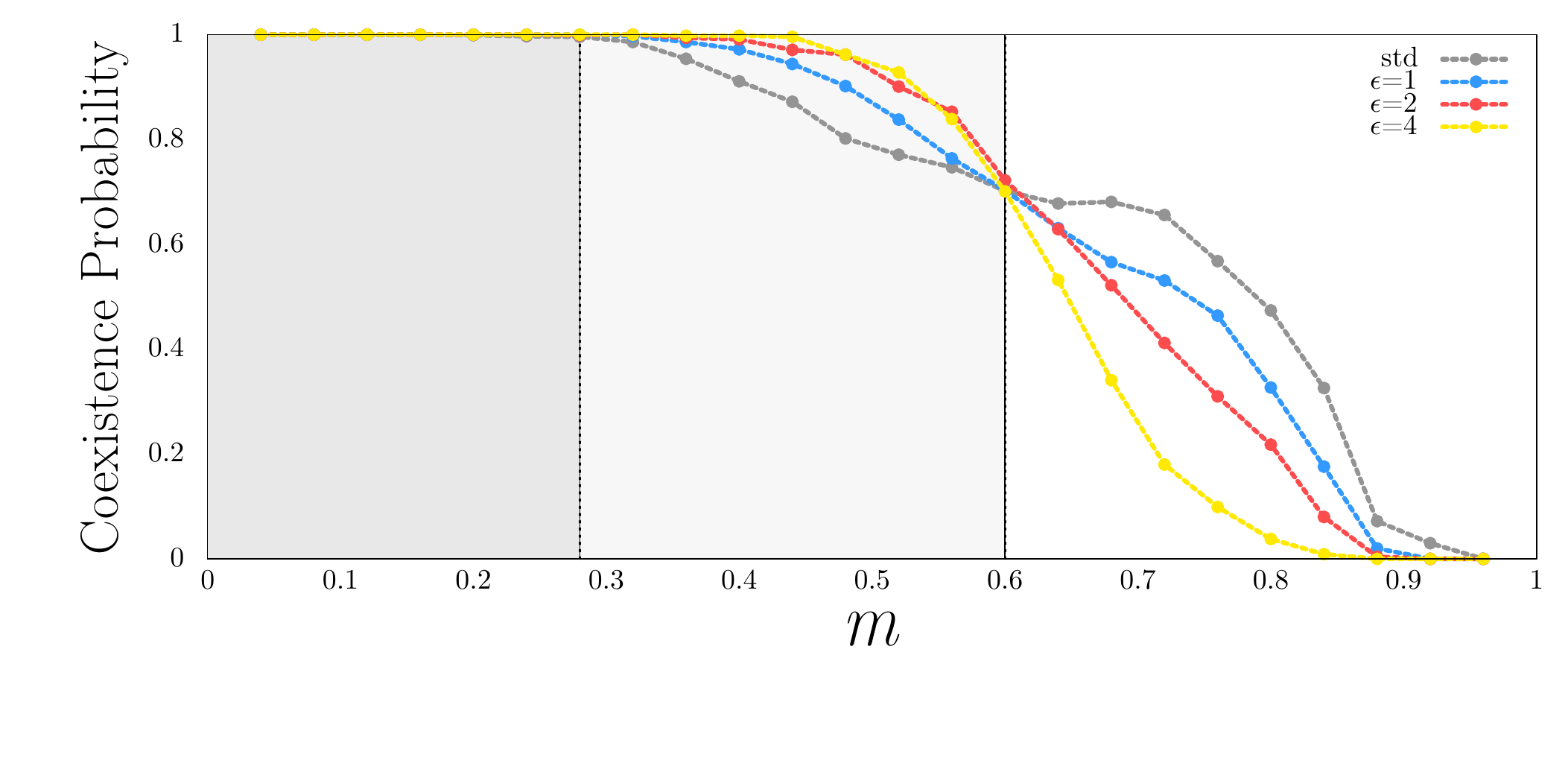}
        \caption{}\label{fig7b}
    \end{subfigure}\\
     \begin{subfigure}{.49\textwidth}
        \centering
        \includegraphics[width=78mm]{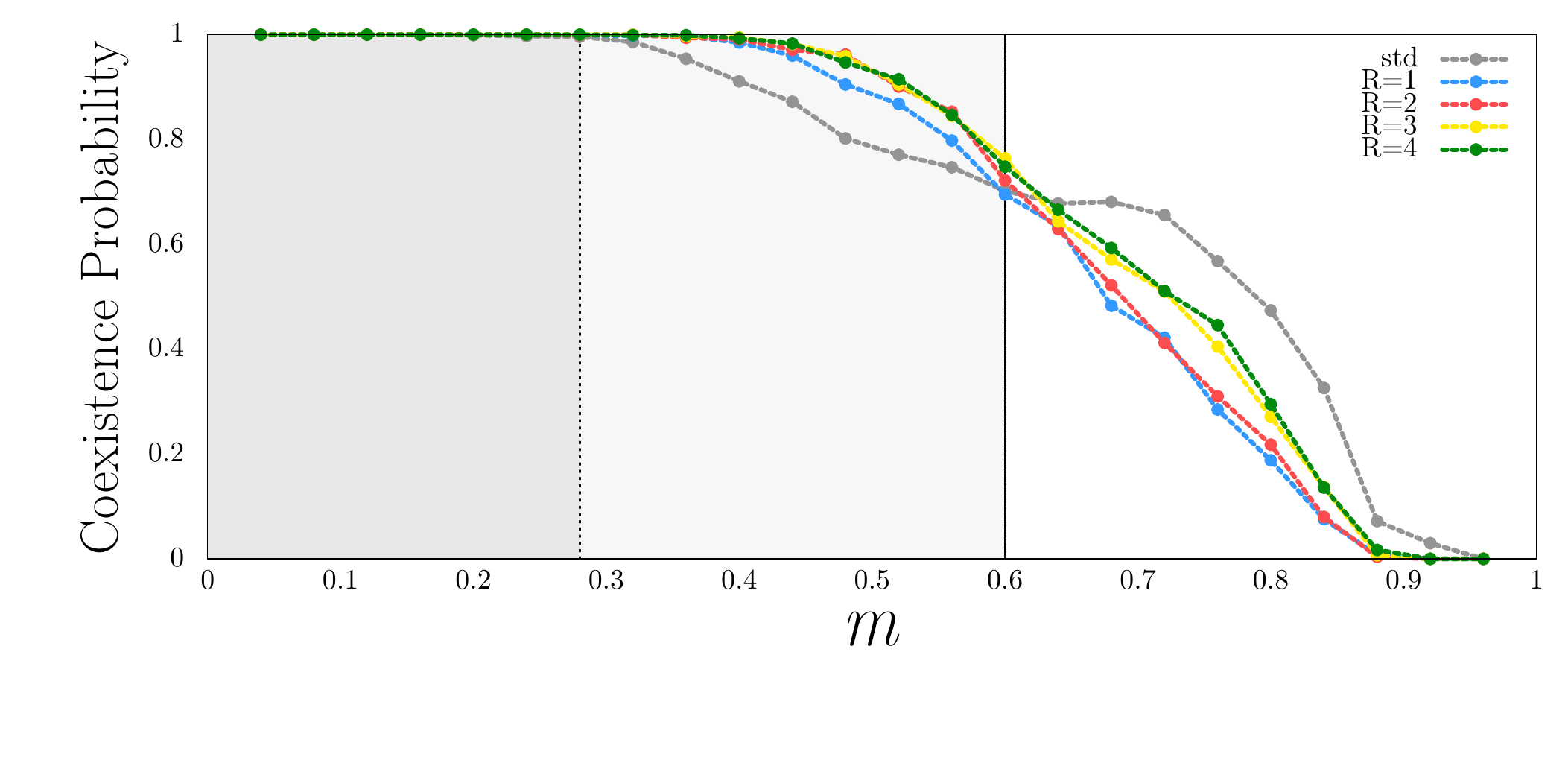}
        \caption{}\label{fig7c}
    \end{subfigure}
 \caption{
Coexistence probability as a function of the mobility $m$.
Figure \ref{fig7a}, ~\ref{fig7b}, and ~\ref{fig7c} show the outcomes for various $\alpha$, $\epsilon$, and $R$, respectively. The grey line shows the coexistence probability for the standard model. Low, intermediate, and high mobility regimes are identified by dark grey, light grey, and white backgrounds.}
  \label{fig7}
\end{figure}
Finally, we aim to quantify the interference of adaptive movement strategy in biodiversity maintenance. This investigation was implemented running $1000$ simulations in lattices with $100^2$ grid points for $ 0.04\,<\,m\,<\,0.96$ in intervals of $ \Delta\, m\, =\,0.04$, for several values of $\alpha$, $\epsilon$, and $R$. The simulations started from different random initial conditions and ran for a timespan of $100^2$ generations. Coexistence occurs only if at least one individual of all species is present at the end of the simulation, $I_i (t=10000) \neq 0$ with $i=1,2,3$. If at least one species is absent, the simulation results in extinction. The coexistence probability is defined as the fraction of implementations resulting in coexistence. The selection and reproduction probabilities were set to be $s\,=\,r\,=\,(1-m)/2$. 

The outcomes depicted in Fig. \ref{fig7a} were taken for 
various conditioning factors: $\alpha=0.3$ (blue line), $\alpha=0.6$ (yellow line), and $\alpha=1.0$ (red line) - the results were obtained for $\epsilon=2$ and $R=2$. In Fig. \ref{fig7b} appear the results for several responsiveness factors: $\epsilon=1$ (blue line), $\epsilon=2$ (red line), and $\epsilon=4$ (yellow line) - the results were obtained for $\alpha=1.0$ and $R=2$. Finally, Fig. \ref{fig7c} depicts the coexistence probability for organisms with different perception radius: $R=1$ (blue line), $R=2$ (red line), $R=3$ (yellow line), and $R=4$ (green line) - the simulations ran for $\alpha=1.0$ and $\epsilon=2$.
In Figs. \ref{fig7a}, \ref{fig7b}, and \ref{fig7c}, the grey line shows the coexistence probability for the standard model, where no organism intelligently adjusts the mobility.

For better analysing our results, we separated the mobility probabilities into three intervals: i) low mobility: $m\,\leq \,0.28$ (dark grey background); ii) intermediate mobility: $0.28\,<\, m\,\leq \,0.60$ (light grey background); iii) high mobility: $m\,\geq\,0.60$ (white background). 

\subsection{Low mobility}

The adaptive movement behaviour does not jeopardise biodiversity for $m\,\leq \,0.28$. This happens because the oscillations on the species densities are not sufficient to break the system stability, as observed in the examples of Figs.~\ref{fig6a}, ~\ref{fig6d}, and ~\ref{fig6g} for $m=0.2$.  

\subsection{Intermediate mobility, $\alpha=1.0$}

For $0.28\,< m \leq 0.60$, the behavioural strategy promotes biodiversity if all organisms are conditioned to move adaptively ($\alpha=1.0$), independent of 
$\epsilon$ and $R$. In this scenario, if $\epsilon\,>\,1$ or $R\,>\,1$, coexistence is more probable, according to Figs.~\ref{fig7b} and ~\ref{fig7c} - coloured lines compared with the grey one. This happens because, for intermediate mobility probabilities, the oscillations in species densities are more prominent for small $\epsilon$ and $R$, as presented in Figs.~\ref{fig6e} and ~\ref{fig6h}. We discover that biodiversity is more benefited if the fraction of organisms following the environment cues grows; besides, the coexistence probability is higher in the case where organisms can scan a larger area and adjust their mobility abruptly.

\subsection{Intermediate mobility, $\alpha<1.0$}
In the case of not all individuals being prepared to move strategically, the adaptive movement promotes coexistence for a shorter range of intermediate mobility. According to Fig.~\ref{fig7a}, the coexistence probability increases for $0.28\,<\,m\,\leq \,0.52$, when compared with the standard model, with the results being improved for larger $\alpha$. 

\subsection{High mobility}
The outcomes for $m \geq 0.6$ prove that biodiversity is jeopardised if at least $60\%$ of individuals learn to adjust their mobility in response to local stimuli. In this scenario, 
even if the organism's responsiveness is minimal or its ability to scan the environment is limited, species coexistence will be harmed. This is indicated by the coloured lines in Figs.~\ref{fig7b} and ~\ref{fig7c}, which also reveals that coexistence probability is more endangered for large $\epsilon$ and small $R$. This happens because species densities' oscillations increase with $\epsilon$ and decrease with $R$, as in the case of $m=0.8$ in Figs.~\ref{fig6f} and ~\ref{fig6i}. Moreover, the outcomes for the particular case of $\alpha=0.3$ indicate a slight improvement in the coexistence probability; thus, we conclude that 
if the fraction of conditioned organisms is small, biodiversity may be promoted even if organisms move with high mobility probability.

\section{Discussion and Conclusions}
\label{sec7}
We studied the spatial rock-paper-scissors game, where interactions among individuals are selection, reproduction, and mobility. In our stochastic simulations, organisms of one species can calculate the level of attractiveness of their neighbourhood, thus deciding to accelerate in case of being in danger or decelerating when in their comfort zone. We calculate the changes in the species segregation and species densities considering that not all organisms are physically or cognitively prepared to adjust their movement to environmental cues, for organisms with
different responsiveness strength and perception radius.

Our outcomes reveal that the larger the proportion of individuals ready to react to environmental cues, the more significant are the consequences on the spatial patterns and population dynamics. In agreement with recent results of the effects of the adaptive movement strategy in rock-paper-scissors models \cite{Adap}, the more significant the proportion of individuals conditioned to move adaptively, the shorter the characteristic length of typical spatial domain occupied by the species is; consequently, the species population declines.
This happens because species $1$ proliferates as organisms of species $2$ are rapidly eliminated. Although the increase in the short-term fitness, this process is disadvantageous in the long-term population size: the population decline of species $2$
diminishes the chances of organisms of species $1$ conquering more territory. 
This is similar to the spreading of a high virulent disease whose severity causes the host death, weakening the disease transmission; the same happens when an aggressive predator overexploits the local prey population causing a long-term difficulty to reproduce \cite{localfitness}.

However, the unevenness in the territorial control caused by the adaptive motor response is not bad for biodiversity, as reported in studies of other behavioural movement strategies (\cite{Moura,howdirectional}).
Our discoveries unveiled that, for low mobility range $m\,\leq \,0.28$, it does make any difference if individuals of every species move randomly or one out of them evolve to move adaptively. However, more interesting results were reached for intermediate mobility range ($0.28\,<\,m\,\leq \,0.60$): the chances of species coexist does increase if all organisms can control their mobility rate in response to the environment attractiveness. More, we found that the coexistence probability increases 
if organisms can perceive longer distances and respond powerfully to local stimuli.  
Conversely, adaptive movement strategy endangers biodiversity for high mobility ($m\,> \,0.60$), irrespective of the organisms' skill to perform the behavioural strategy.

In summary, in cyclic models, the adaptive movement behaviour is not advantageous for the species at the population level since it leads to a spatial density decrease. Nevertheless, this may promote biodiversity in case of individuals do not move at high mobility rates.

\acknowledgments
We thank CNPq, ECT, Fapern, and IBED for financial and technical support.


\begin{thebibliography}{10}
\expandafter\ifx\csname url\endcsname\relax\def\url#1{\texttt{#1}}\fi


\bibitem{Motivation1}
\Name{Fraenkel G. S. \and Gunn D. L.} \REVIEW{The American Naturalist }{75}{761}{1941}{604}.

\bibitem{kinesisnew1}
\Name{Gorban A. N. \and Çabukoǧlu N.} \REVIEW{Ecological Complexity}{33}{2018}{75-83}.


\bibitem{adaptive1}
\Name{Riotte-Lambert L. \and Matthiopoulos J.} \REVIEW{Trends in Ecology \& Evolution}{35}{2020}{163-174}.

\bibitem{adaptive2}
\Name{Abrams P. A.} \REVIEW{The American Naturalist}{169}{2007}{581–594}.

\bibitem{Dispersal}
\Name{Bonte D. \and Dahirel M.} \REVIEW{Oikos}{126}{2017}{472-479}.

\bibitem{BENHAMOU1989375}
\Name{Benhamou S. \and Bovet P.} \REVIEW{Animal Behaviour}{38}{3}{1989}{375–383}.

\bibitem{coping}
\Name{Martin J. \and Benhamou S. \and Yoganand K. \and Owen-Smith N.} \REVIEW{PLOS ONE}{10}{2015}{e0118461}.

\bibitem{NEW}
\Name{Gilbert S. F.} \REVIEW{Evol. Dev.}{14}{2012}{20-8}.



\bibitem{foraging}
\Name{Abrams P. A.} \REVIEW{The American Naturalist }{124}{1}{1984}{80-96}.


\bibitem{BUCHHOLZ2007401}
\Name{Buchholz R. \and Hector A.} \REVIEW{Trends in Ecology \& Evolution}{22}{2007}{401}.


\bibitem{repellent}
\Name{Pettersson H. \and Amundin M. \and Laska M. } \REVIEW{Frontiers in Behavioral Neuroscience}{12}{2018}{152}.

\bibitem{Causes}
\Name{Bowler D. E. \and Benton T. G.} \REVIEW{Biol Rev. Camb. Philos. Soc.}{80}{2005}{205-225}.

\bibitem{MovementProfitable}
\Name{Barraquand F. \and Benhamou S.} \REVIEW{Ecology}{89}{2008}{3336–3348}.




\bibitem{animats}
\Name{Maes P. \and Mataric M. J. \and Meyer J. A. \and Pollack J. \and Wilson S. W.} \REVIEW{From Animals to Animats 4: Proceedings of the Fourth International Conference on Simulation of Adaptive Behavior}{1996}{55-64}.


\bibitem{Coli}
\Name{Kerr B. \and Riley M. A. \and Feldman M. W. \and Bohannan B. J. M.} \REVIEW{Nature}{418}{2002}{171}.

\bibitem{bacteria}
\Name{Kirkup B. C. \and Riley M. A.} \REVIEW{Nature}{428}{2004}{412}.

\bibitem{Allelopathy}
\Name{Durret R. \and Levin S.} \REVIEW{J. Theor. Biol.}{185}{1997}{165}.


\bibitem{lizards}
\Name{Sinervo B. \and Lively C. M.} \REVIEW{Nature}{380}{1996}{240}.

\bibitem{Extra1}
\Name{Volkov I. \and Banavar J. R. \and Hubbell S. P. \and Maritan A.} \REVIEW{Nature}{450}{2007}{45}.

\bibitem{mutation0}
\Name{Park, J. and Do, Y. and Huang, Z.-G. and Lai, Y-.C.} \REVIEW{Chaos}{23}{2013}{023128}.

\bibitem{mutation1}
\Name{Park, J.} \REVIEW{Europhysics Letters}{126}{2019}{38004}.


\bibitem{mutation2}
\Name{Mohd, M. H. and Park, J.} \REVIEW{Chaos, Solitons \& Fractals}{153}{2021}{111579}.

 \bibitem{Reichenbach-N-448-1046}
\Name{Reichenbach T. \and Mobilia M \and Frey E.} \REVIEW{Nature}{448}{2007}{1046}.

\bibitem{Moura}
\Name{Moura B. \and Menezes J.} \REVIEW{Scientific Reports}{11}{2021}{6413}.

\bibitem{howdirectional}
\Name{Avelino P. P. \and Bazeia D. \and Losano L. \and Menezes J. \and de Oliveira B. F. \and Santos M. A.} \REVIEW{Phys. Rev. E}{97}{2018}{032415}.



\bibitem{Anti1}
\Name{Menezes J.} \REVIEW{Phys. Rev. E}{103}{2021}{052216}.

\bibitem{Anti2}
\Name{Menezes J. \and Moura B.} \REVIEW{Phys. Rev. E}{104}{2021}{054201}.

\bibitem{Agg}
\Name{Menezes J. \and Rangel E. \and Moura B.} \REVIEW{Ecological Informatics}{69}{2022}{101606}.

\bibitem{Adap}
\Name{Tenorio M. \and Rangel E. \and Menezes J.} \REVIEW{ArXiv}{2203.06531}{2022}.

\bibitem{doi:10.1002/ece3.4446}
\Name{Revynthi A. M. \and Egas M. \and, Janssen A. \and Sabelis M. W.} \REVIEW{Ecology and Evolution}{8}{21}{2018}{10384–10394}.

\bibitem{SabelisII}
\Name{van Wijkm M., \and De Bruijn P. J. A. \and Sabelis M. W.} \REVIEW{J. Chem. Ecol.}{34}{2008}{791–803}.

\bibitem{uneven}
\Name{Menezes J. \and Moura B. \and Pereira T. A.} \REVIEW{Europhysics Letters}{126}{2019}{18003}.


\bibitem{leonard}
\Name{May R. M. \and Leonard W. J.} \REVIEW{SIAM J. Appl. Math.}{29}{1975}{243–253}.

\bibitem{localfitness}
\Name{Rauch, E. M. and Sayama, H. and Bar-Yam, Y.} \REVIEW{Phys. Rev. Lett.}{88}{22}{2002}{228101}.






\end{thebibliography}
\end{document}